\documentstyle[preprint,aps]{revtex}

\begin{document}
\title{Inelastic Coulomb scattering rate of a multisubband Q1D electron gas}
\author{M. Tavares and G.-Q. Hai}
\address{Instituto de F\'{\i }sica de S\~{a}o Carlos, \\
Universidade de S\~{a}o Paulo, 13560-970 S\~{a}o Carlos, SP, Brasil.}
\maketitle

\begin{abstract}
In this work, the Coulomb scattering lifetimes of electrons in two coupled
quantum wires have been studied by calculating the quasiparticle self-energy
within a multisubband model of quasi-one-dimensional (Q1D) electron system.
We consider two strongly coupled quantum wires with two occupied subbands.
The intrasubband and intersubband inelastic scattering rates are caculated
for electrons in different subbands. Contributions of the intrasubband,
intersubband plasmon excitations, as well as the quasiparticle excitations
are investigated. Our results shows that the plasmon exictations of the
first subband are the most important scattering mechanism for electrons in
both subbands.
\end{abstract}


\section{Introduction}

Recently single-particle properties of electrons in quasi-one-dimensional
(Q1D) systems have attracted considerable interest. By calculating the
quasiparticle renormalization factor $Z_{k}$ \cite{mhbook} and the momentum
distribution function $n_{k}$ around the Fermi surface, Hu and Das Sarma 
\cite{sarmagrande} have clarified that a clean Q1D electron system in a
semiconductor quantum wire shows the Luttinger liquid behavior, whereas even
slightest amount of impurities restores the Fermi surface and the
Fermi-liquid behavior remains. Within an one-band model, they calculated the
self-energy due to electron-electron Coulomb interaction for an unclean Q1D
system by using the leading-order GW dynamical screening approximation.\cite
{sarmagrande,quinn} This self-energy gives rise to the inelastic Coulomb
scattering rate which plays a fundamental role in measurement of the
quasiparticle lifetime in Q1D doped systems. In the inelastic-scattering
processes, the quasiparticle loses energy each time it scatters and its
lifetime provides information about the different excitation channels its
energy relaxation occurred through.\cite{zheng,muitos}

This paper is devoted to study the collective excitation and the Coulomb
inelastic-scattering rate in a multisubband Q1D doped semiconductor. The
numerical calculations are applied to two strongly coupled GaAs/AlGaAs
quantum wires in which the occupation of higher subbands provides more
scattering channels. We show that the intersubband coupling and intersubband
excitations can be important in the energy relaxation process.

\section{Theoretical formulation}

By assuming zero thickness in the $z$ direction, the subband energies $E_{n}$
and the wave functions $\phi _{k,n}(y)$ are obtained from the numerical
solution of the coupled one-dimensional Schr\"{o}dinger equation in the $y$
direction. The Coulomb inelastic-scattering rate of an electron in a subband 
$n$ with momentum $k$ is obtained \cite{mhbook} by the imaginary part of the
screened exchange self-energy $\Sigma _{n}\left[ k,\xi _{n}(k)\right] $,
where $\xi _{n}\left( k\right) =E_{n}+\hbar ^{2}k^{2}/2m^{\ast }-\mu $ is
the single-particle energy with\ $\mu $ being the chemical potential and $%
m^{\ast }$ the electron effective mass in GaAs. At zero temperature \cite
{vinter}, the screened exchange self-energy is given by 
\[
\Sigma _{n}\left[ k,\xi _{n}(k)\right] =\frac{i}{(2\pi )^{2}}\int dq\int
d\omega ^{\prime }\times 
\]
\begin{equation}
\sum_{n_{1}}V_{nn_{1}n_{1}n}^{s}(q,\omega ^{\prime })G_{n_{1}}^{(0)}\left(
k+q,\xi _{n}(k)-\omega ^{\prime }\right) ,  \label{self1}
\end{equation}
where $G_{n_{1}}^{(0)}(k,\omega )$ is the bare Green's function of
noninteracting electrons and $V_{nn_{1}n_{1}n}^{s}(q,\omega ^{\prime })$ is
the dynamically screened electron-electron interaction potential. The
self-energy $\Sigma _{n}\left[ k,\xi _{n}(k)\right] $ was written by taking
the leading pertubative term in an expansion in the dynamically screened RPA
exchange interaction (GW approximation).

Similarly to the one-band model\cite{kuang}, the self-energy in Eq.(\ref
{self1}) can be separated into the frequency-independent exchange and the
correlation part, $\Sigma _{n}\left[ k,\xi _{n}(k)\right] =\Sigma
_{n}^{ex}(k)+\Sigma _{n}^{cor}\left[ k,\xi _{n}(k)\right] .$ The exchange
part is given by 
\begin{equation}
\Sigma _{n}^{ex}(k)=-\frac{1}{2\pi }\int
dq\sum_{n_{1}}V_{nn_{1}n_{1}n}^{b}(q)\;f_{n_{1}}\left( \xi
_{n_{1}}(k+q)\right) ,  \label{exnua}
\end{equation}
where $f_{n}\left( \xi _{n}(k)\right) $ is the Fermi-Dirac distribution
function and $V_{nn_{1}n_{1}n}^{b}(q)$ is the multisubband electron-electron
bare interaction. Notice that $\Sigma {}_{n}^{ex}(k)$ is real because the
bare interaction potential $V_{nn_{1}n_{1}n}^{b}(q)$ is totally real.
Therefore, one only needs to analyze the imaginary part of $\Sigma _{n}^{cor}%
\left[ k,\xi _{n}(k)\right] $, since it gives rise to the imaginary part of
the self-energy which we are interested in. After some algebra, we find that
the Coulomb inelastic-scattering rate for an electron in a subband $n$ with
momentum $k$ is given by 
\[
\sigma _{n}(k)=\sum_{n^{\prime }}\sigma _{n,n^{\prime }}(k)=-%
\mathop{\rm Im}%
\Sigma {}_{n}^{cor}\left[ k,\xi _{n}(k)\right] , 
\]
with 
\[
\sigma _{n,n^{\prime }}(k)=\frac{1}{2\pi }\int dq\times 
\]
\[
\left\{ 
\mathop{\rm Im}%
\left[ V_{nn^{\prime }n^{\prime }n}^{s}\left( q,\xi _{n^{\prime }}(k+q)-\xi
_{n}(k)\right) \right] -V_{nn^{\prime }n^{\prime }n}^{b}(q)\right\} 
\]
\begin{equation}
\times \left\{ \theta \left( \xi _{n}(k)-\xi _{n^{\prime }}(k+q)\right)
-\theta \left( -\xi _{n^{\prime }}(k+q)\right) \right\} ,  \label{sigma}
\end{equation}
where $\theta \left( x\right) $ is the standard step function. In the above
equation, the frequency integration has already been carried out, since the
bare Green's function $G_{n_{1}}^{(0)}$ can be written as a delta function
of $\omega $.

\section{Results and discussions}

In this section, we will present some numerical results of $\sigma
_{n,n^{\prime }}(k)$ within a two-subband model ($n,n^{\prime }=1,2$). In
order to understand the inelastic Coulomb scattering processes of the Q1D
electron gas, we will firstly analyze the collective excitation dispersion
relation in this multisubband system to show the different channels of
energy relaxation.

We consider two strongly double symmetric GaAs/AlGaAs quantum wires of
widths $l_{w_{1}}=l_{w_{2}}=150$ \AA\ and barrier height $V_{0}=228$ meV.
Between them there is a potential barrier of width $\ l_{b}=10$ \AA . The
total density of electrons is taken as $N_{e}=1.0$ $a_{b}^{\ast -1}$, where $%
a_{b}^{\ast }=\kappa \hbar ^{2}/2m^{\ast }e^{2}$ is the effective Born
radius with $\kappa $ being the dielectric constant of the static lattice
and $e$ the electron charge. In such a system, the electron densities in the
first $(n=1)$ and the second $(n=2)$ subbands are $0.73$ $a_{b}^{\ast -1}$
and $0.27$ $a_{b}^{\ast -1}$, respectively. Fig. 1 shows the collective
excitation dispersion relation of the Q1D electron gas. We find two
intrasubband collective excitation modes indicated by $(1,1)$ and $(2,2)$
and two intersubband modes indicated by $(1,2)$ and $(1,2)^{\prime }$. The
intrasubband plasmon dispersions are approximately linear in the
long-wavelength limit, while the intersubband ones have finite energy values
at $q=0$. The low-energy intrasubband plasmon is mainly due to the second
subband, while the high-energy one is mainly due to the first subband. We
also see a quite large depolarization shift of the intersubband plasmon mode 
$(1,2)$. The shadow areas indicated by $QPE_{nn^{\prime }}$ present the
quasiparticle excitation regions which can result in Landau damping of the
collective excitation modes. Notice that, due to the symmetry of the system,
the intrasubband and intersubband modes do not couple to each other in such
a way that the intersubband quasiparticle excitations do not damp the
intrasubband plasmon modes and vice versa.\cite{nosso,hng} The existence of
two undamped intersubband collective modes is a particular feature of the
Q1D system with two occupied subbands. The occupation of the second subband
opens up a gap inside the inter-subband quasiparticle excitations $QPE_{12}$
region where the intersubband mode $(1,2)^{\prime }$ appears.

In Fig. 2, we show the Coulomb inelastic-scattering rate $\sigma
_{nn^{\prime }}(k)$ as a function of the electron wave vector $k$ in this
two subband coupled quantum wire system. The parameters of the system are
the same as in Fig. (1). The solid curves denote the intrasubband scattering
rate for which the initial and final states of the electron are in the same
subband. The dotted ones correspond to the intersubband scattering rate.
Fig. 2(a) shows the scattering rate for an electron initially in the first
conduction subband. The numerical results show that the intrasubband
scattering rate $\sigma _{nn}(k)=0$ at $k=k_{F1}$ and $k_{F2}$ ( $%
k_{F1}=1.15a_{b}^{\ast -1}$ and $k_{F2}=0.42a_{b}^{\ast -1}$ ) as it should
be. The scattering rate $\sigma _{11}(k)$ presents a very pronounced peak at
about $k=1.9$ $a_{b}^{\ast -1}>k_{F1}$ corresponding to the contribution
from the plasmon excitation of the first subband. Such a contribution opens
up a channel through which the electron energy can relax by emitting one
plasmon. This strong peak is due to the onset of the scattering of the
plasmon mode $(1,1),$ restricted by the conservation of energy and momentum,
joining up with the divergency of the density of states at the bottom of the
1D subband. We also observe the contributions from the quasiparticle
excitation of the second subband at small wave vectors ($k<k_{F2}$), and
from the plasmon excitation of the second subband, peaked at $k\simeq
0.5a_{b}^{\ast -1}$. The intersubband scattering rate $\sigma _{12}(k)$
(dotted line) has contributions from the intersubband plasmon excitation
(high-energy), at $k\simeq 3.35$ $a_{b}^{\ast -1}$, and from quasiparticle
intersubband excitation (region $QPE_{12}$), at small values of wave
vectors. The scattering rate $\sigma _{12}(k)$ is much smaller than\ $\sigma
_{11}(k)$ because the corresponding intersubband transition is from lower to
higher subband and, moreover, intersubband plasmon has higher energy. The
sum of the two curves in Fig. 2(a) yields the total scattering rate for an
electron in the first subband. Fig. 2(b) shows the scattering rate for
electrons in the second subband. The intrasubband scattering rate $\sigma
_{22}(k)$ (solid curve) also shows a pronounced peak at $k\simeq 1.9$ $%
a_{b}^{\ast -1}$ corresponding to the scattering of the intrasubband plasmon
mode $(1,1),$ while the scattering from the plasmon mode $(2,2)$ of the
second subband only leads to a small peak at $k\simeq 0.7$ $a_{b}^{\ast -1}$%
. There is a small shoulder between the first and the second subband plasmon
excitation. This shoulder denotes the contribution of the intrasubband
quasiparticle excitation (region $QPE_{11}$) to the $\sigma _{22}(k)$. The
strong scattering of the plasmon $(1,1)$ of the first subband to the $\sigma
_{22}(k)$ indicates a strong coupling of the in-plane motion of the
collective excitations from the different subbands. From $\sigma _{22}(k)$
and $\sigma _{11}(k)$, we also observe a relatively large contribution from $%
QPE_{11}$ and $QPE_{22}$, respectively. This is because the virtue
intersubband scattering (coupling) makes the energy-momentum conservation
less restrictive in the scattering processes. The dotted curve in Fig. 2(b)
shows the contribution from the intersubband plasmon excitation. One sees
that the intersubband scattering $\sigma _{21}(k)$ has a peak located at
lower momentum $k$ comparing to that of $\sigma _{12}(k)$. In the scattering
process from the second to the first subband, the electron transfer its
potential energy to kinetic energy which facilitates the scattering.

Our results show that the plasmon excitations of the first subband are the
most important scattering mechanism for electrons in both subbands. This is
so because: (i) the strong intersubband coupling and (ii) there are more
occupied states in the first subband on which the self-energy effects are
stronger and the scattering becomes more pronounced.

\section{Summary}

In summary, we have calculated the multisubband collective excitations and
the Coulomb inelastic-scattering rate of a quasi-one-dimensional electron
gas in two strongly coupled quantum wires with two occupied subbands. We
have analyzed the different modes of collective excitations and the
corresponding scattering channels through which the electrons, in the
different subbands, lose their energies. The role of each excitation mode in
the inelastic-scattering relaxation process is clarified. We find that the
plasmons in the lowest subband are responsible for the most important
scattering channel even for the electrons in higher subbands. The
intersubband plasmon modes open up more scattering channels, but their
scattering rates are about one order of magnitude smaller than that of the
main scattering mechanism.

\section{Acknowledgments}

This work is supported by FAPESP (Funda\c{c}\~{a}o de Amparo \`{a} Pesquisa
do Estado de S\~{a}o Paulo) and CNPq (Conselho Nacional de Desenvovimento
Cient\'{i}fico e Tecnol\'{o}gico).

\begin{figure}[tbp]
\caption{ The multisubband collective excitation spectra for two strongly coupled 
symmetric $GaAs/Al_{0.3}Ga_{0.7}As$ quantum wires of widths 
$l_{w_{1}}=l_{w_{2}}=150$ \AA\ and barrier height $V_{0}=228$ meV. 
The two quantum wires are separated by a barrier of $l_{b}=10$ \AA. 
The electron density is $N_e =1.0 a_b^{*-1}$. The plasmon 
dispersions (dotted lines) are indicated by $(n,n')$. The shadow areas present 
the quasiparticle excitation regions indicated by $QPE_{nn'}$.}
\end{figure}

\begin{figure}[tbp]
\caption{ The Coulomb inelastic-scattering rates in the coupled quantum wires
for an electron initially in (a) the first subband and (b) the second subband. 
Solid (dotted) curves give the intrasubband (intersubband) scattering rates. 
The parameters are the same as in Fig.(1). }
\end{figure}

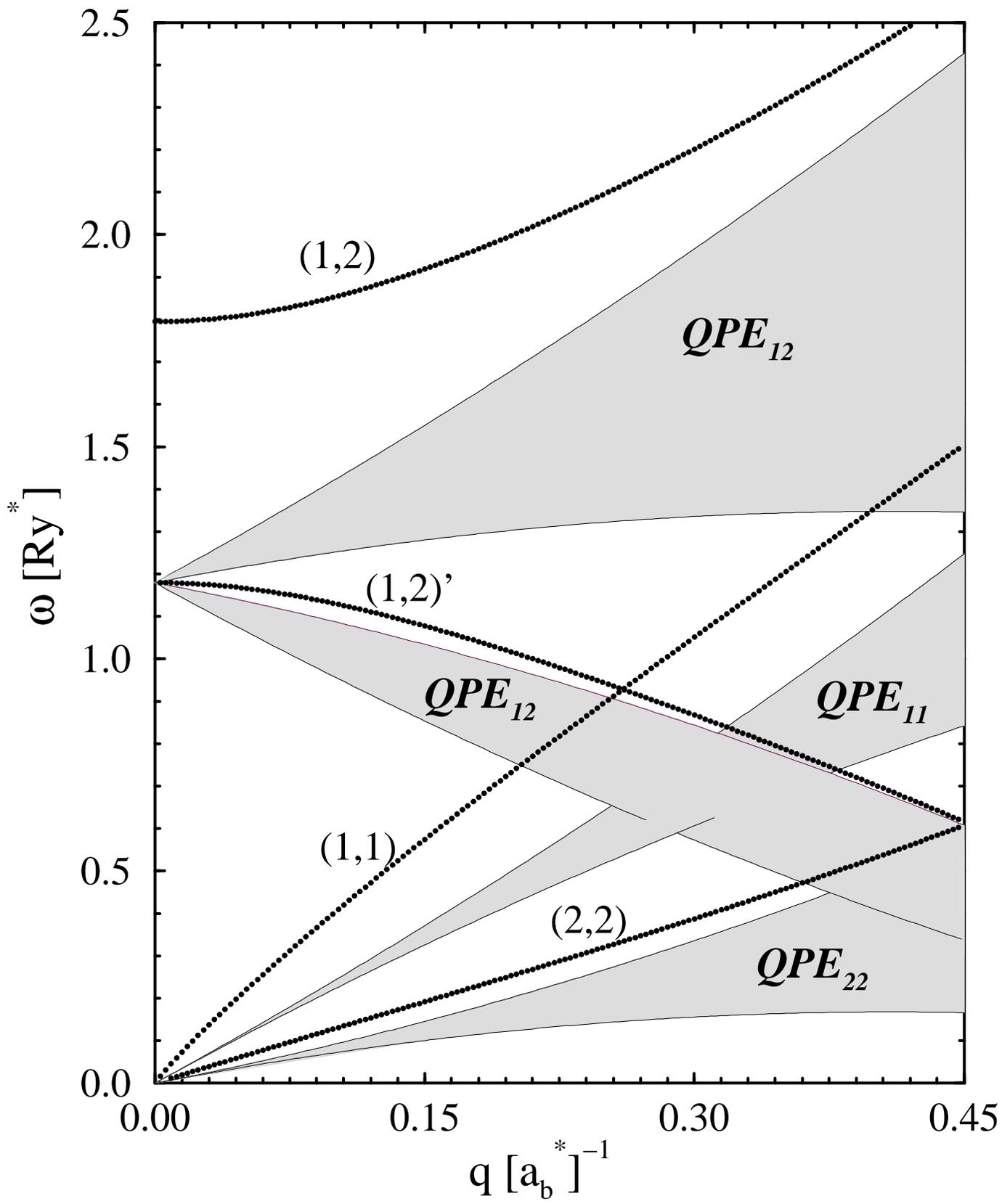
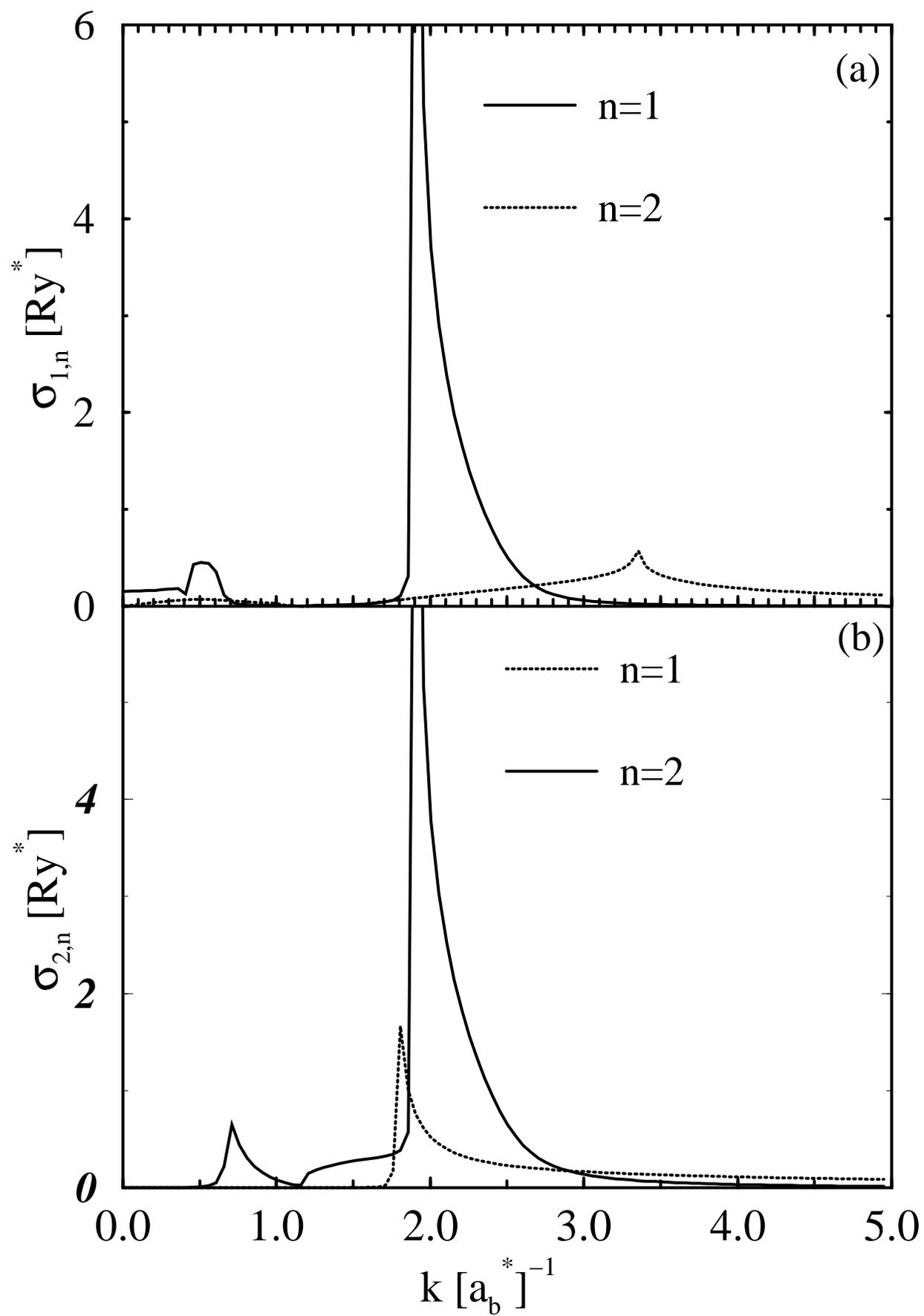
\end{document}